\documentclass[11pt]{article}
\usepackage{psfig}

\begin{document}

\title{A New Free Core Nutation Model with Variable Amplitude and Period}
\author{Zinovy Malkin \\
 \it Institute of Applied Astronomy RAS, St.~Petersburg, Russia}
\date{February 4, 2004}
\maketitle

\begin{abstract}
Three most long and dense VLBI nutation series obtained at the Goddard
Space Flight Center, Institute of Applied Astronomy, and U.S. Naval
Observatory were used for investigation
of the Free Core Nutation (FCN) contribution to the celestial pole offset.
Some recent studies have showed that the FCN period or/and phase does not
remain constant, but varies in a rather wide range of about 410--490
days (for equivalent period).
To implement this result in the practice,
a new FCN model with variable amplitude and period (phase)
is developed. Comparison of this model with observations shows better
agreement than existing one.  After correction of the differences
between observed VLBI nutation series and the IAU2000A model, they decreased
to the level about 100~$\mu$as.
\end{abstract}

\section{Introduction}

Free Core Nutation (FCN) of the Earth is predicted more than a century ago as
a common rotational mode of a body having an ellipsoidal solid
shell and fluid core.
Investigation of the FCN is an important scientific task.
First, the FCN parameters determined from observations provide
valuable, sometimes unique, information about processes in the
Earth's interior.  From the practical point of view, accurate modelling
the FCN term, including prediction, is necessary to compute celestial
pole offset with accuracy compatible with modern VLBI observations.

The IAU2000A model
based on the MHB2000 model developed in (Mathews, {\it et~al.}, 2002),
and adopted as a new IAU standard
can predict a regular part of the nutation with accuracy of about 100~$\mu$as.
However, the FCN contribution is much larger, up to 400~$\mu$as, which
yields degradation of accuracy in modelling celestial pole offset,
if FCN not accounted.  It is well known also that the FCN
contribution gives the prevailing contribution to the power spectrum
of the differences between observed nutation series and modern models.

The IAU2000A model, like the previous ones,
is constructed as a Poison series, and does not include free mode terms,
such as the FCN, which cannot be presented as a Poison series term
with predictable parameters.  For this reason, FCN parameters are to
be determined from the VLBI observations.

Historically, the FCN frequency is considered as a constant fundamental
value included in transfer function expression describing the relationship
between the amplitudes of nutation terms for real and rigid Earth.
Many authors have made an effort to estimate the FCN period and possible
reasons for its excitation (see {\it e.g.} Mathews and Shapiro, 1995;
Brzezi\'nski and Petrov, 1998; Shirai and Fukushima, 2001a;
Herring {\it et al.}, 2002; Malkin and Terentev, 2003a, 2003b).
They found the FCN period to be in the range of 425--435 solar days,
with average value about $430^d$.

Recently, it was found from a wavelet analysis of VLBI nutation
series that the FCN period likely varies in a range of about 410--490 days
(Malkin and Terentev, 2003a, 2003b).
This result is also confirmed by means of another method, Short-time
Periodogram with Gabor Function, proposed by T.~Shirai (Shirai {\it et al.},
2004).  Of course, found variability of the FCN period may be fully or
partially an transformation of the variations of the FCN phase, which
may has more geophysical meaning.  However, geophysical considerations
of the FCN variability lie beyond of this study.

In this paper we develop an practical model for computation of the
FCN contribution to the celestial pole offset taken into account
variability both amplitude and period (phase) of the FCN oscillation.

\section{Computation of the FCN parameters}

Three most long and dense VLBI nutation series obtained at the Goddard
Space Flight Center (GSF), Institute of Applied Astronomy (IAA), and U.S. Naval
Observatory (USN), each containing more than 3000
estimates of the nutation angles for the period from 1979 up to now
were used for investigation of the FCN parameters.

Firstly, estimates of $d \psi$ and $d \varepsilon$ w.r.t. the
IAU1976/1980 nutation model computed at the GSF and USN were transformed to
the $d  X_c$ and $d  Y_c$ w.r.t. IAU2000A model (IAA series already
contains this data).

\begin{figure}[ht]
\centering
\hbox{
\psfig{figure=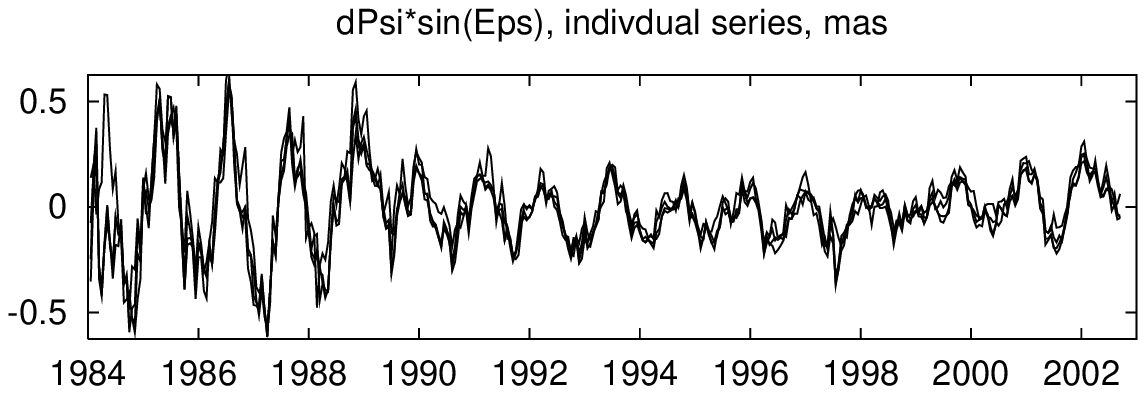,width=0.5\textwidth,clip=}
\psfig{figure=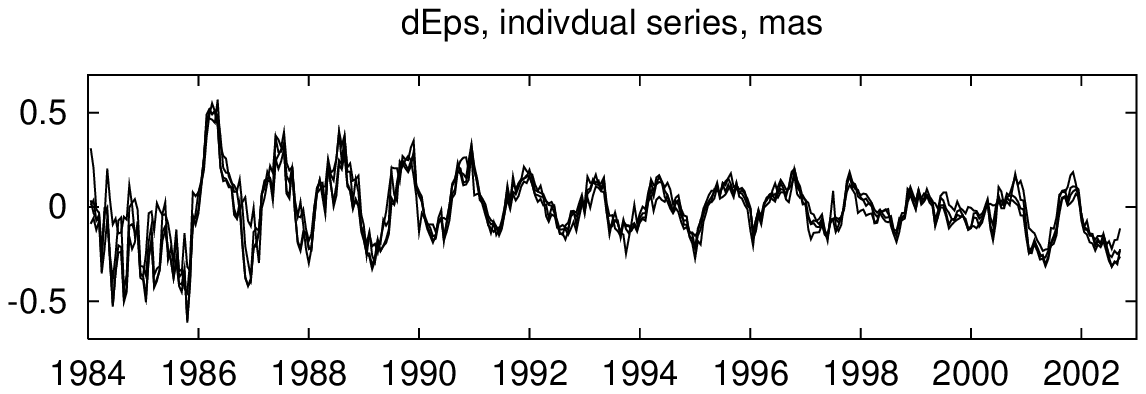,width=0.5\textwidth,clip=}
}
\hbox{
\psfig{figure=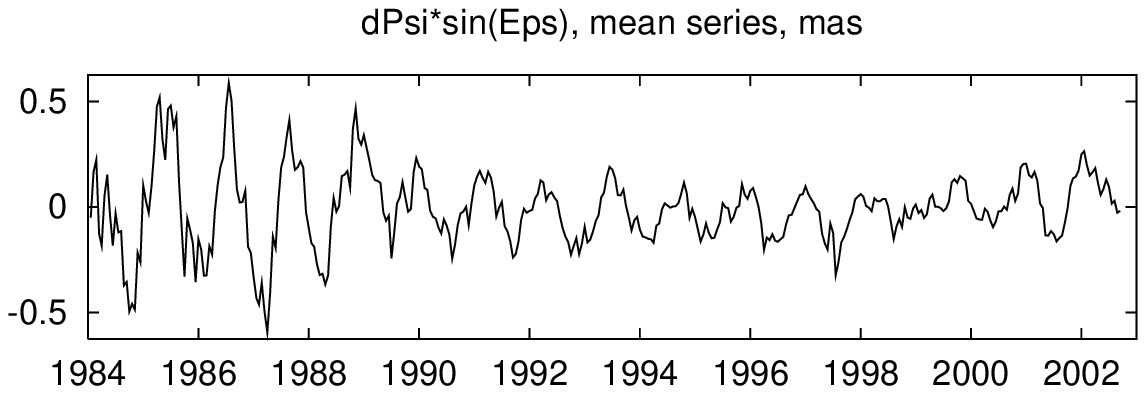,width=0.5\textwidth,clip=}
\psfig{figure=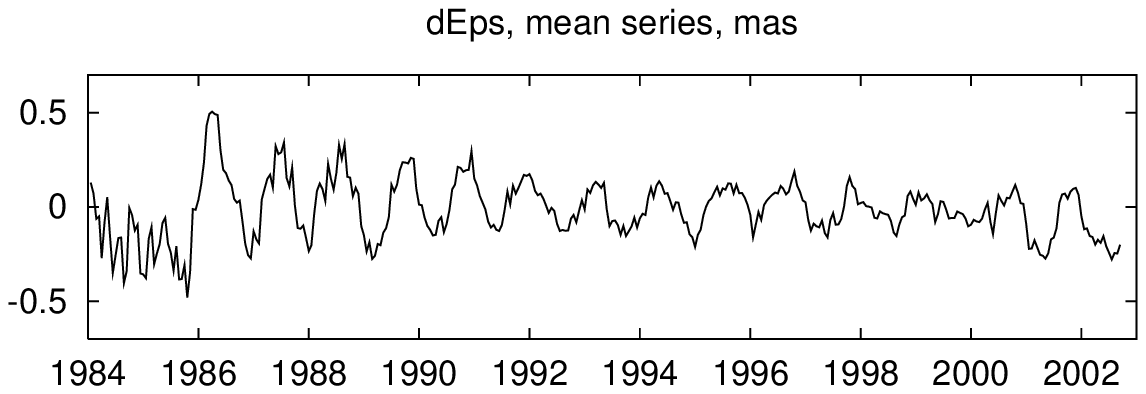,width=0.5\textwidth,clip=}
}
\caption{Smoothed differences between observed nutation series and the
IAU2000A model.}
\label{fig:diff_fcn}
\end{figure}

\begin{figure}[ht]
\centering
\psfig{figure=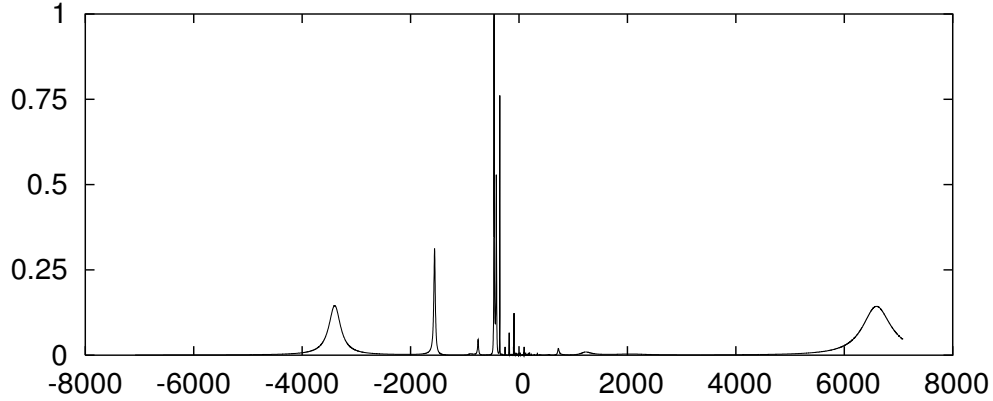,width=0.85\textwidth,clip=}
\caption{Spectrum of the differences between observed nutation series and
the IAU2000A model.}
\label{fig:diff_spectrum}
\end{figure}

Then combined series was computed.  Since preliminary analysis showed
that three input series are of very similar quality, no weighting was
applied, however formal errors reported in the input series were scaled
for uniformity.  After that, input series were averaged, saving only
epochs present in all the series.
Band-pass Gaussian filtering was applied to the combined series.
Transfer function of the filter at the FCN frequency is 0.988.
Figure~\ref{fig:diff_fcn} shows obtained smoothed differences
between observed nutation series and the IAU2000A model,
and spectrum of the differences is presented in Figure~\ref{fig:diff_spectrum},

At this step we also computed the smoothed values at equally sampled epochs
with 20-day step.  However, a smoothed series given at the original epochs
also can be used for analysis with a similar final result
(Malkin and Terentev, 2003a, 2003b).

The FCN amplitude time series was computed using the simple formula
\begin{equation}
A(t) = \sqrt{(d X_c(t))^2+(d Y_c(t))^2} \,,
\end{equation}

Indeed, using such an approach we
suppose that all differences can be
attributed to the FCN, but this seems to be a good approximation
to reality.  However, a resonance impact on the nutation terms at the
frequencies close to the FCN one, evidently should be accounted for
in future developments.

Finally, the FCN period variation was computed using a wavelet analysis
as described in (Malkin and Terentev, 2003a, 2003b).

\section{Computation of the FCN contribution}

Let us consider how a model with variable amplitude and period (phase)
can be used in practice.  We can describe the FCN term as
\begin{equation}
\begin{array}{rcl}
d X_c & = & A(t)\sin(\Phi(t)) \,,\\
d Y_c & = & A(t)\cos(\Phi(t)) \,.
\end{array}
\end{equation}

Mathematically (not geophysically, indeed!),
we can suppose three equivalent models for the FCN phase $\Phi(t)$
\begin{equation}
\Phi(t) = \left\{
\begin{array}{l}
\displaystyle\frac{2\pi}{P(t)}\;t + \Phi_{\scriptscriptstyle 0} \,, \\[1.5em]
\displaystyle\frac{2\pi}{P_{\scriptscriptstyle 0}}\:t + \Phi(t) \,, \\[1.5em]
\displaystyle\frac{2\pi}{P(t)}\,t + \Phi(t) \,,
\end{array}
\right.
\end{equation}
where P is the FCN period, and zero subscripts mean constant values.
In other words, we can consider three models:
with variable period and constant phase, variable phase and constant period,
or variable both period and phase.  Of course, this is a subject of geophysical
consideration, but doesn not matter for an empiric FCN model using
time variations of the FCN parameters found from analysis of the observed data.

In practice, one can compute $\Phi(t)$ as
\begin{equation}
\Phi(t) = \int\limits_{t_0}^t {\displaystyle\frac{2\pi}{P(t)}\,dt} +
\varphi_{\scriptscriptstyle 0}\,,
\end{equation}
where $\varphi_{\scriptscriptstyle 0}$ is the parameter to be adjusted.

\begin{figure}[ht]
\centering
\hbox{
\psfig{figure=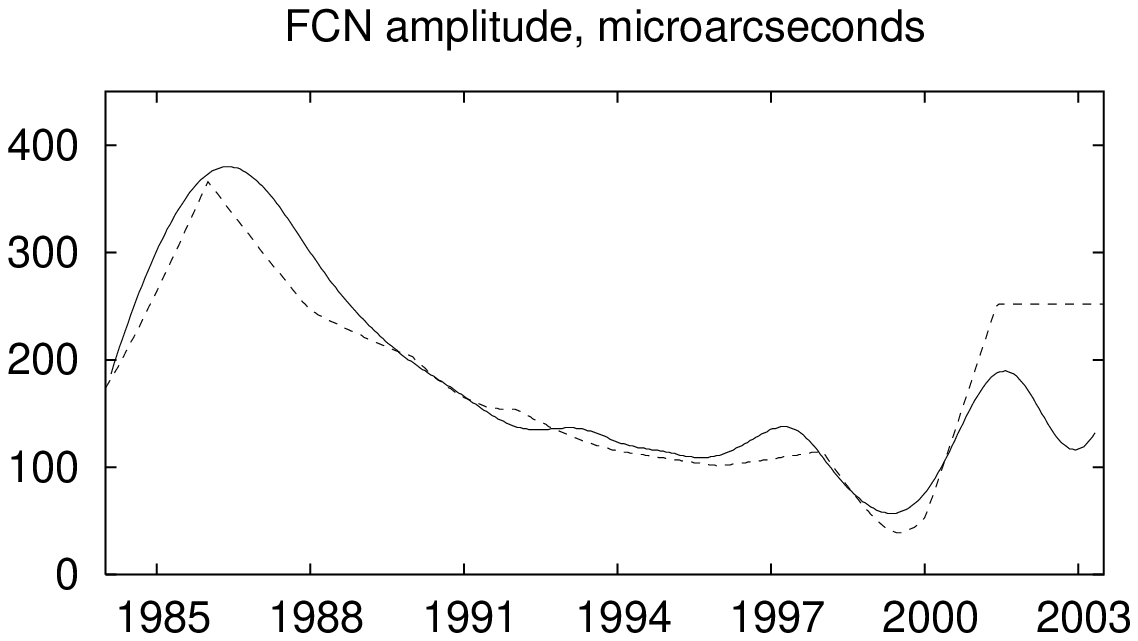,width=0.5\textwidth,clip=}
\psfig{figure=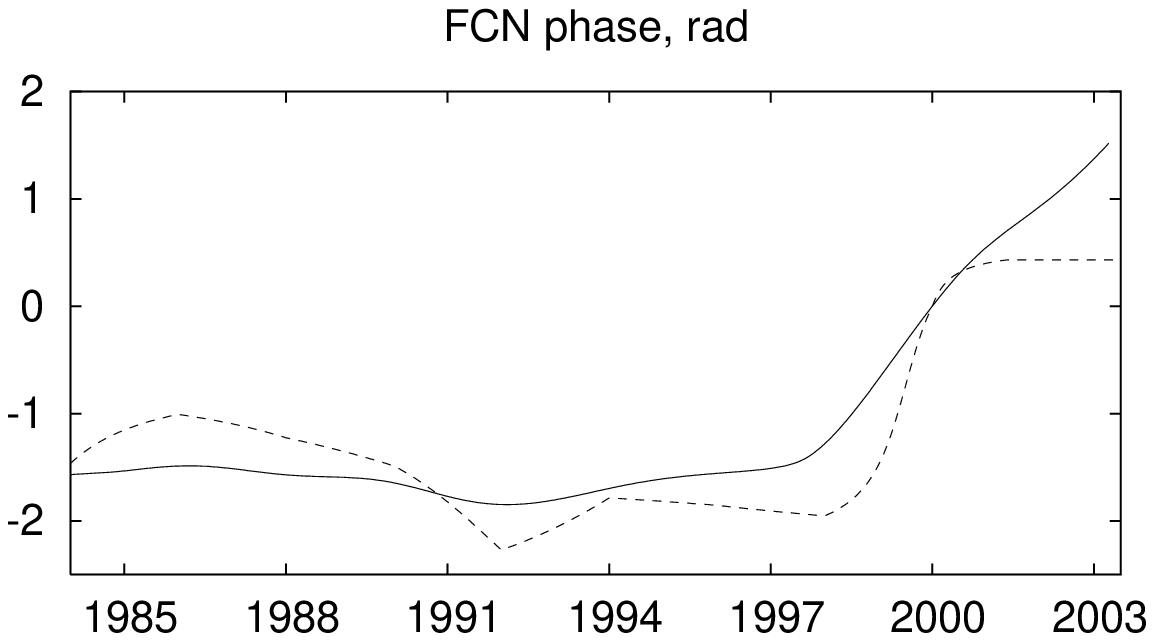,width=0.5\textwidth,clip=}
}
\caption{The FCN amplitude and phase variations found in this study (solid
line), and a comparison with the MHB2000 model (dashed line).}
\label{fig:fcn_ap_mhb}
\end{figure}

Variations of the FCN amplitude $P(t)$ and phase $\Phi(t)$ are
shown in Figure~\ref{fig:fcn_ap_mhb} along with the corresponding
FCN parameters included in the MHB2000 model which is, in fact,
also a model with variable phase and amplitude, though this is not
stated explicitly (we used the text of the FCN\_NUT routine included
in the MHB\_2000 code to extract the FCN(MHB) amplitude and phase variations).
One can see that both models show similar behavior of the FCN parameters,
however new approach allow us to get continues, non-inflecting
and predictable functions $A(t)$ and $\Phi(t)$.
Comparing these two models one should keep
in mind that MHB2000 model is developed only for the period till 2001.4,
and after this epoch the difference between the models grows rapidly.

\begin{figure}[ht]
\centering
\hbox{
\psfig{figure=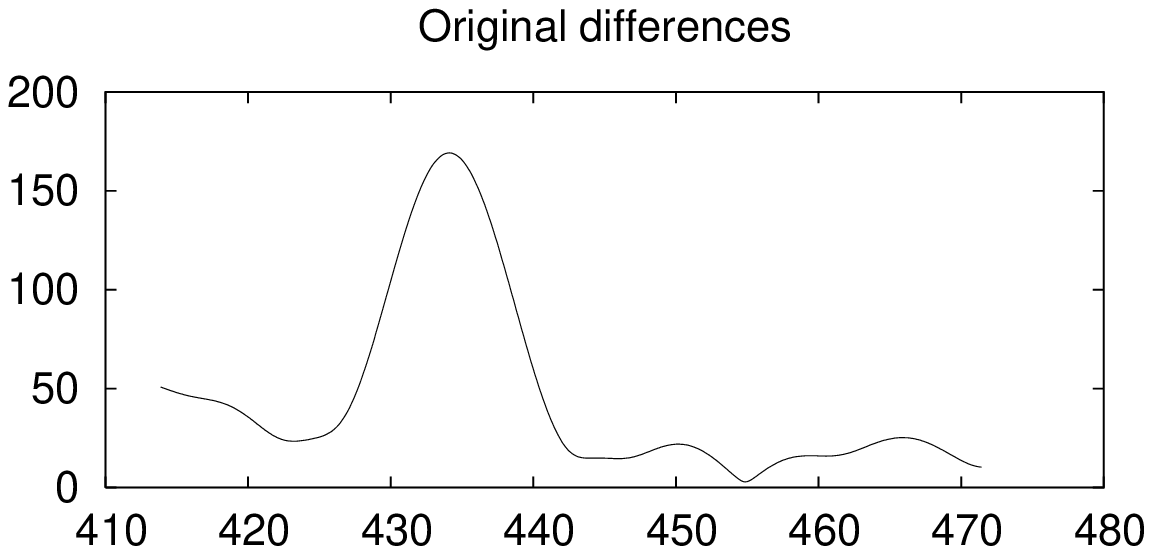,width=0.5\textwidth,clip=}
\psfig{figure=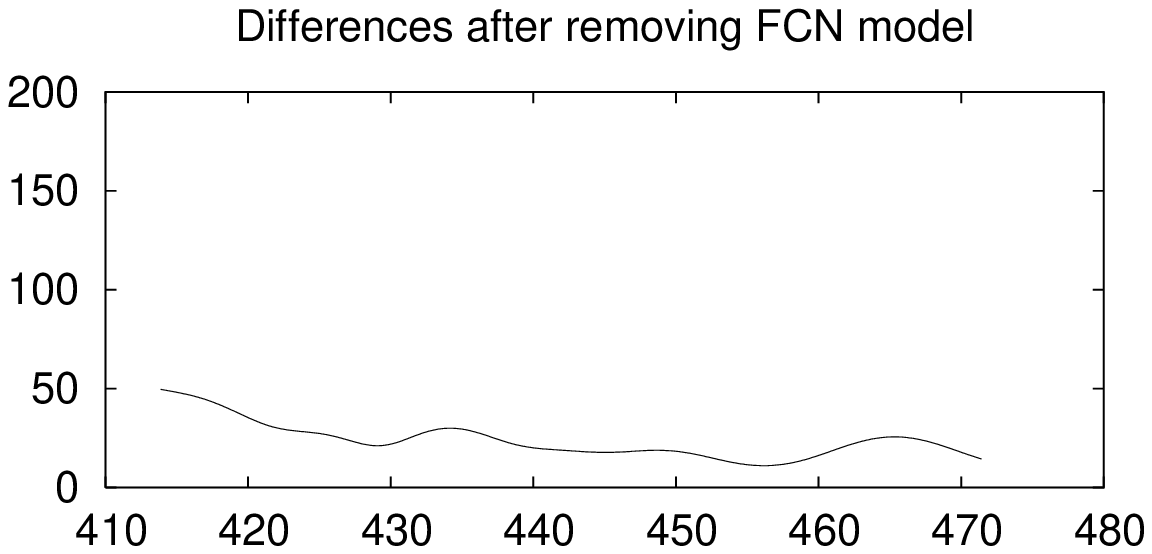,width=0.5\textwidth,clip=}
}
\caption{Spectrum of the differences between observed nutation series and
the IAU2000A model, period in days, amplitude in~$\mu$as.}
\label{fig:fcn_spectrum}
\end{figure}

Figure~\ref{fig:fcn_spectrum} shows spectra of the differences
between observed nutation series and the IAU2000A model computed
for raw differences and after removing the FCN contribution.
One can see that the FCN signal is completely eliminated.

However, the differences between observed nutation series and
model have a noise of various origins with the rms compatible
with the FCN contribution.  To estimate the actual contribution
of the FCN model to this noise we have computed rms of differences
between the observations and the IAU2000A model after
applying three different FCN models:
no FCN (raw differences), extracting the MHB2000 FCN term,
and extracting the FCN term according to the model described here.
The results are shown in Table~\ref{tab:wrms_fcn}.

\begin{table}[ht]
\centering
\caption{WRMS of differences with two FCN models, $\mu$as
(No~-- no FCN model, MHB~-- MHB2000 FCN model, New~-- model
proposed in this paper applied; NEOS~-- NEOS-A VLBI sessions
observed in 1993--2001,
R1R4~-- IVS R1 and R4 VLBI sessions only observed since 2002).}
\label{tab:wrms_fcn}
\tabcolsep=3pt
\begin{tabular}{cccccccccc}
\hline
Series & \multicolumn{3}{c}{All sessions} & \multicolumn{3}{c}{NEOS}
       & \multicolumn{3}{c}{R1R4} \\
\cline{2-10}
       &  \multicolumn{3}{c}{FCN model} & \multicolumn{3}{c}{FCN model}
       &  \multicolumn{3}{c}{FCN model} \\
       & No & MHB & New & No & MHB & New & No & MHB & New \\
\hline
GSF  & 166 & 146 & 138 & 138 & 122 & 120 & 134 & 150 & 102 \\
IAA  & 170 & 152 & 144 & 140 & 123 & 123 & 138 & 154 & 111 \\
USN  & 161 & 144 & 136 & 138 & 122 & 122 & 136 & 156 & 107 \\
\hline
Mean & 156 & 136 & 126 & 131 & 113 & 112 & 129 & 146 & ~97 \\
\hline
\end{tabular}
\end{table}

One can see that accounting for the FCN contribution leads to
decreasing of the differences.
Especially interesting is the last part of the table corresponding
to the period of observations 2002--2003.  Using the MHB2000 FCN model
for this period leads to degradation of differences between
observations and the IAU2000A model, which is natural for this
model is developed only for epochs untill 2001.4.

\begin{figure}[ht]
\centering
\hbox{
\psfig{figure=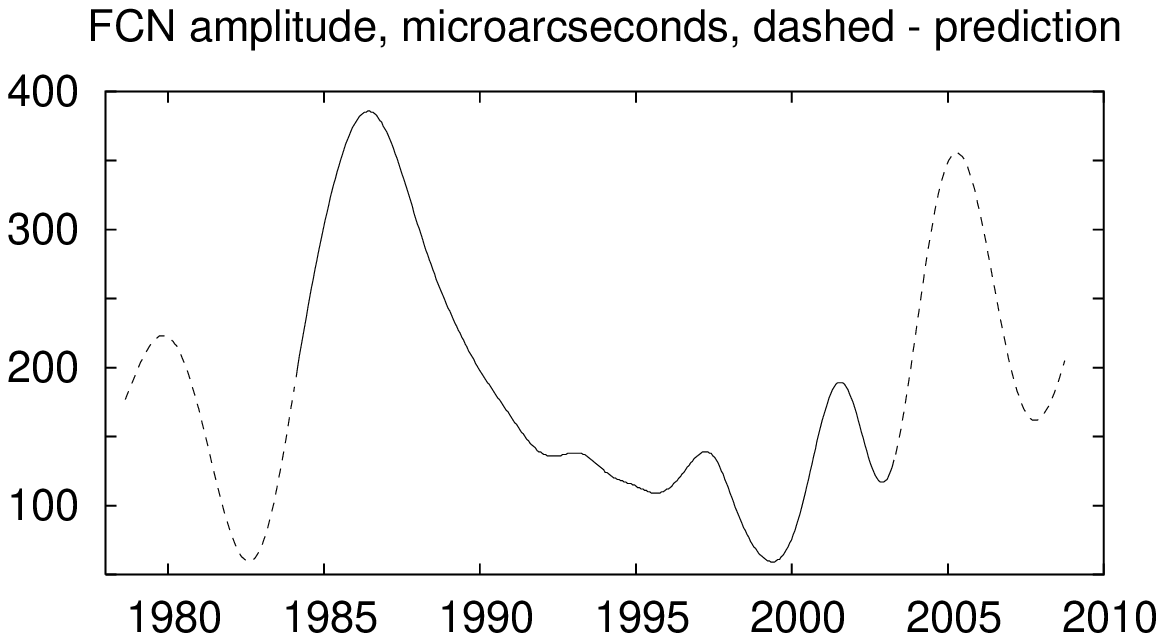,width=0.5\textwidth,clip=}
\psfig{figure=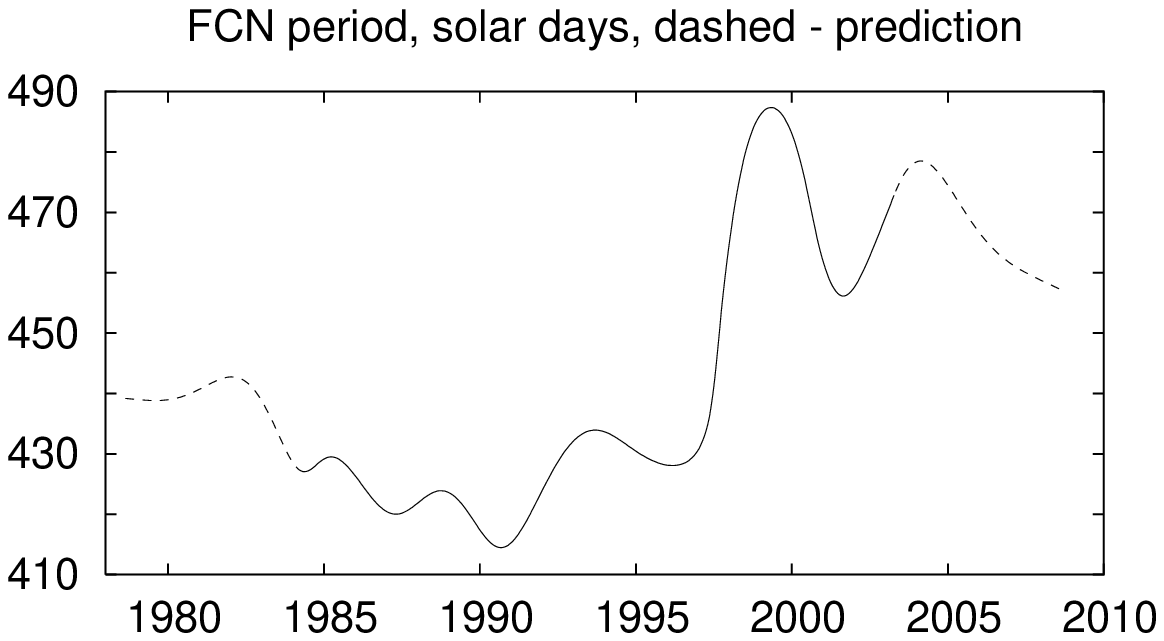,width=0.5\textwidth,clip=}
}
\caption{Examples of predictions of the FCN amplitude and phase.}
\label{fig:fcn_pred}
\end{figure}

A FCN model with variable period and phase allow us to try a new approach
to FCN prediction.  One can consider two possibilities.  The first one
is a prediction of actual FCN contribution, which is developed {\it e.g.}
in (Brzezi\'nski and Kosek, 2004).  Another possibility is to predict
functions $A(t)$ and $\Phi(t)$ separately, and then use predictions
to construct the FCN contribution using the formulas given above.
Figure~\ref{fig:fcn_pred} presents a variant of such a prediction
obtained using ARMA method. It is interesting to compare both
approaches to a FCN prediction in details.  Please note that in our
final computation we replaced observed series for the period 1979
with predicted one which seems to be more accurate taking into
account relatively large errors in the VLBI observations made before 1984.

\section{Conclusions}

We have developed a new FCN model with variable amplitude and period (phase)
which provides a computation of
a continuous FCN contribution to the celestial pole offset with
good accuracy for whole interval of the VLBI observations, and
convenient prediction of the FCN contribution.
Using this model allow us to reduce the differences
between VLBI observations and model to the level 100$\mu$as.

It is clear that the proposed model is a pure empiric one.  Considerable
efforts should be made to understand the physical origin of the
variability of the FCN period and/or phase, and its consequences on
a theory of nutation.

The proposed model is routinely used in the VLBI processing at the
Institute of Applied Astronomy since September 2003.

\end{document}